\documentclass[twocolumn,showkeys,prd,nofootinbib,floatfix]{revtex4-1}

\usepackage{amsfonts,amsmath,amssymb}
\usepackage{graphicx,graphics}

\def\mnras{Mon. Not. Roy. Astron. Soc.}

\def\mnras{MNRAS}
\def\apj{ApJ}
\def\apjl{ApJ}

\def\prd{Phys. Rev. D}

\def\prl{Phys. Rev. Lett.}

\begin{document}

\title{THE QUANTUM EMISSION OF AN ALIVE BLACK HOLE\\ \textit{Essay written for the Gravity Research Foundation 2021 awards for essays on Gravitation}}

\author{J.~A.~Rueda$^{1,2,3,4,5}$}
\email{jorge.rueda@icra.it (corresponding author)}
\author{R.~Ruffini$^{1,2,6}$}
\email{ruffini@icra.it}

\affiliation{$^1$ICRANet, Piazza della Repubblica 10, I--65122 Pescara, Italy}
\affiliation{$^2$ICRA, Dipartimento di Fisica, Sapienza Universit\`a di Roma, P.le Aldo Moro 5, I--00185 Rome, Italy}
\affiliation{$^3$ICRANet-Ferrara, Dipartimento di Fisica e Scienze della Terra, Universit\`a degli Studi di Ferrara, Via Saragat 1, I--44122 Ferrara, Italy}
\affiliation{$^4$Dipartimento di Fisica e Scienze della Terra, Universit\`a degli Studi di Ferrara, Via Saragat 1, I--44122 Ferrara, Italy}
\affiliation{$^5$INAF, Istituto de Astrofisica e Planetologia Spaziali, Via Fosso del Cavaliere 100, 00133 Rome, Italy
}
\affiliation{$^6$INAF, Viale del Parco Mellini 84, 00136 Rome  Italy}

\date{\today}

\begin{abstract}

A long march of fifty years of successive theoretical progress and new physics discovered using observations of gamma-ray bursts, has finally led to the formulation of an efficient mechanism able to extract the rotational energy of a Kerr black hole to power these most energetic astrophysical sources and active galactic nuclei. We here present the salient features of this long-sought mechanism, based on gravito-electrodynamics, and which represents an authentic shift of paradigm of black holes as forever ``\textit{alive}'' astrophysical objects.

\end{abstract}
\pacs{Valid PACS appear here}

\maketitle

\section{Introduction}\label{sec:1}

Traditionally, rotating black holes (BHs) have been described by the Kerr \citep{1963PhRvL..11..237K} and the Kerr-Newman (for nonzero charge) metrics \citep{1965JMP.....6..918N} which adopt the spacetime to fulfill: (i) matter vacuum, (ii) asymptotic flatness, and (iii) global stationarity. These conditions led to the primordial view of BHs either as ``\textit{dead}'' objects or as sinks of energy. Subsequently, it was realized that BHs, much as the thermodynamical systems, may interact with their surroundings leading to reversible and irreversible transformations \citep{1970PhRvL..25.1596C, 1971PhRvD...4.3552C}. This result led to the one of the most important concepts in BH physics and astrophysics, i.e. the BH mass-energy formula \citep{1971PhRvD...4.3552C}:
\begin{equation}
\label{eq:Mbh}
M^2 = \left(M_{\rm irr} + \frac{Q^2}{4 G M_{\rm irr}} \right)^2 + \frac{c^2}{G^2}\frac{J^2}{4 M^2_{\rm irr}},
\end{equation}
which relates the BH mass, $M$, to three independent parameters: its irreducible mass, $M_{\rm irr}$, charge, $Q$, and angular momentum, $J$. The expression of the BH mass-energy (\ref{eq:Mbh}) was soon confirmed by \citet{1972CMaPh..25..152H}.

It turns immediately out from Eq.~(\ref{eq:Mbh}) that the BH extractable energy
\begin{equation}
    E_{\rm extr} = (M-M_{\rm irr}) c^2,
\end{equation}
could reach up to $50\%$ of $M c^2$ in a maximally charged BH (charge to mass ratio equal to unity), and up to $29\%$ in a maximally rotating BH (spin to mass ratio equal to unity). This extraordinary result swung the attention of the astrophysics community to the alternative view of ``\textit{alive}'' BHs since their energy could be extracted and be used to power astrophysical sources! This novel view of BHs was shaped in ``Introducing the black hole'' by \citet{Ruffini:1971bza} (see also ``On the energetics of black holes'' by R. Ruffini in ~\cite{DeWitt:1973uma}), and since then it has permeated, for fifty years as of this writing, relativistic astrophysics both theoretically and experimentally.

The most energetic known astrophysical sources, gamma-ray bursts (GRBs) and active galactic nuclei (AGNs), were soon identified as primary candidates to be powered by BHs. GRBs, the most powerful transient objects in the sky, release energies of up to a few $10^{54}$~erg in just a few seconds, which implies that their luminosity in gamma-rays, in the time interval of the event, compares to the luminosity of all the stars of the Universe in our past light-cone! GRBs have been thought to be powered (somehow) by stellar-mass BHs, while AGN, releasing up to $10^{46}$~erg~s$^{-1}$ for billion years are thought to be powered by supermassive BHs.

However, since the theoretical formulation of the BH mass-energy formula and the introduction of the concept of extractable energy, every theoretical effort to find a specific mechanism able to efficiently extract the BH energy has been vanified by the implausibility of their actual realization in nature. An example was the \textit{gedanken} Penrose's process which was shown by the authors to be physically not implementable on the ground of traditional physical considerations \cite{1971NPhS..229..177P}. A new physics was needed!.

We have recently introduced in \cite{2019ApJ...886...82R, 2020EPJC...80..300R}  the BH ``\textit{inner engine}'' to explain the high-energy (in the GeV domain) radiation observed in energetic long GRBs, which efficiently extracts the rotational energy of the newborn Kerr BH via a novel gravito-electrodynamical process occurring at the crossroad between quantum electrodynamics and general relativity. We have been guided by our GRB model based on binary-driven hypernova (BdHN) scenario (see Sec.~\ref{sec:2}), which proportioned us with the main ingredients that such a mechanism should have: a Kerr BH, fully ionized matter, and a magnetic field. We have also shown in \cite{2020EPJC...80..300R} that the same mechanism, duly extrapolated to large BH masses, works as well in AGN and can be a copious source of ultrahigh-energy cosmic rays (UHECRs). We summarize in this note the most important properties of this long-sought BH \textit{inner engine} and how it paves the way to a still novel view of BHs as \textit{forever alive} astrophysical objects.

\section{Binary-driven hypernovae}\label{sec:2}

\begin{figure*}
\centering
\includegraphics[width=0.47\hsize,clip]{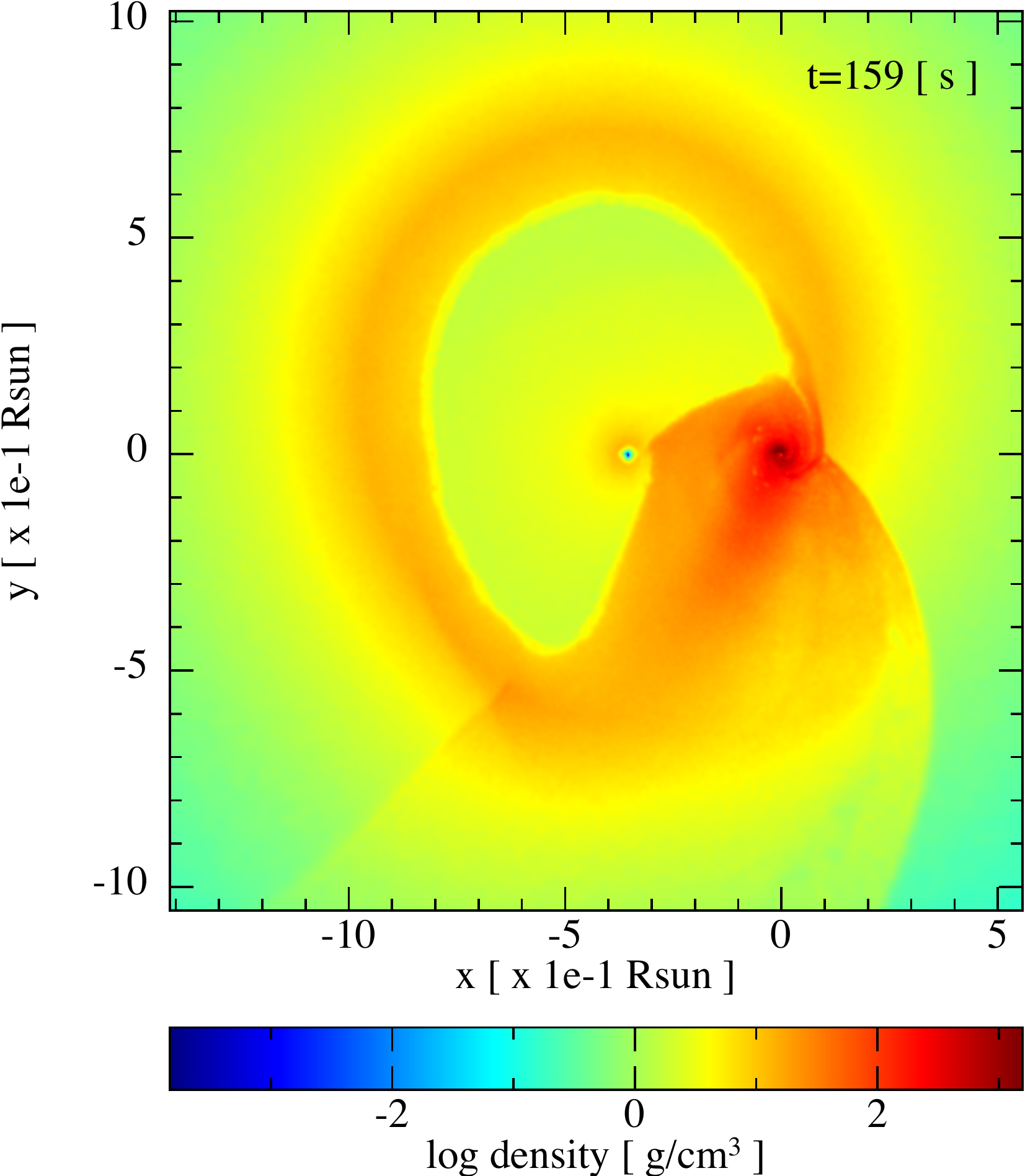}
\includegraphics[width=0.47\hsize,clip]{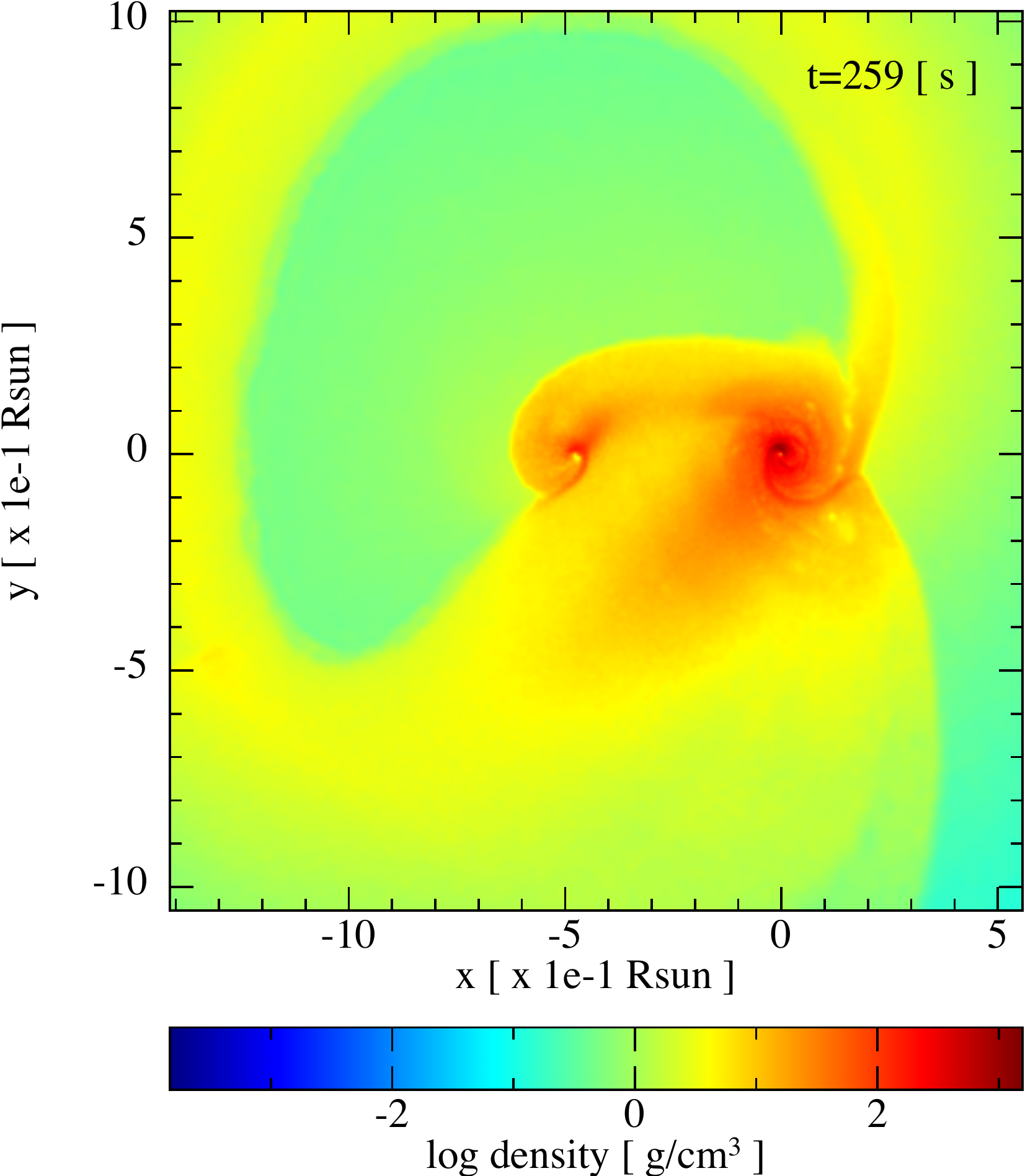}
\caption{Three-dimensional, numerical smoothed particle hydrodynamics (SPH) simulation taken from \citet{2019ApJ...871...14B} of the SN explosion of a CO star in the presence of a binary companion NS. The orbital period is $4.8$~min. The pre-SN CO star mass is $M_{\rm CO}=6.85~M_\odot$ (evolved star from a $25~M_\odot$ zero-age main-sequence progenitor), the $\nu$NS (formed at the center of the SN) mass is $1.85 M_\odot$ and the NS companion mass is $2~M_\odot$. The panels show the mass density on the binary equatorial plane at two selected times from the SN explosion ($t=0$ of the simulation), $159$~s and $259$~s. The reference system is rotated and translated so that the x-axis is along the line that joins the $\nu$NS and the NS, and the axis origin $(0,0)$ is located at the NS position. In this simulation, the NS collapses with a mass $2.26 M_\odot$ and angular momentum $1.24 G M^2_\odot/c$, while the $\nu$NS is stable with mass and angular momentum, respectively, $2.04 M_\odot$ and $1.24 G M^2_\odot/c$.
}
\label{fig:SPHsimulation}
\end{figure*}

Let us start by briefly introducing the BdHN model \cite{2012ApJ...758L...7R, 2014ApJ...793L..36F, 2016ApJ...833..107B, 2019ApJ...871...14B} of long GRBs. The BdHN proposes as GRB progenitor a binary system composed of a carbon-oxygen (CO) star and a neutron star (NS) companion. The gravitational collapse of the iron core of the CO star forms a newborn NS ($\nu$NS) at its center and expels the stellar outermost layers in a supernova/hypernova (SN/HN) explosion. Some of the ejecta fallback onto the $\nu$NS and some other reach the companion, therefore a hypercritical (i.e. highly super-Eddington) accretion process is triggered on both NSs. The accretion onto the $\nu$NS lasts short but is sufficient to spin it up to millisecond rotation rates. For compact binaries (orbital periods $\sim 5$~min), the accretion onto the NS companion makes it to reach, in matter of seconds, the critical mass for gravitational collapse, consequently forming a rotating (Kerr) BH. We have called these long GRBs in which there is BH formation as BdHN of type I (BdHN I) and their isotropic energy release is in the range $10^{53}$--$10^{54}$~erg. Numerical simulations of the above process in one, two and three dimensions have been performed and have confirmed the occurrence of the above succession of physical events \cite{2014ApJ...793L..36F, 2015ApJ...812..100B, 2016ApJ...833..107B, 2019ApJ...871...14B}. Figure~\ref{fig:SPHsimulation} shows an example of three-dimensional simulation taken from \cite{2019ApJ...871...14B}. Since BdHN I keep bound after the explosion, they naturally form NS-BH binaries (see \cite{2015PhRvL.115w1102F} for details).

But not all BdHN form form BHs, up to now, $380$ BdHN I have been identified \cite{2021MNRAS.tmp..868R}. In fact, in binaries with longer orbital periods, e.g. of the order of tens of minutes to hours, the accretion occurs at much lower rates and no BH is formed. The outcome is a BdHN II, a long GRB releasing energies $10^{51}$ to $10^{53}$~erg \citep{2019ApJ...874...39W}, leading to a NS-NS binary. Even less energetics may occur for longer binary periods, we call them BdHN III. The theoretical understanding of BdHN I, II and III have allowed to distinguish their relevant physical processes in the observational data of long GRBs. 

\begin{figure}
\centering
\includegraphics[width=\hsize,clip]{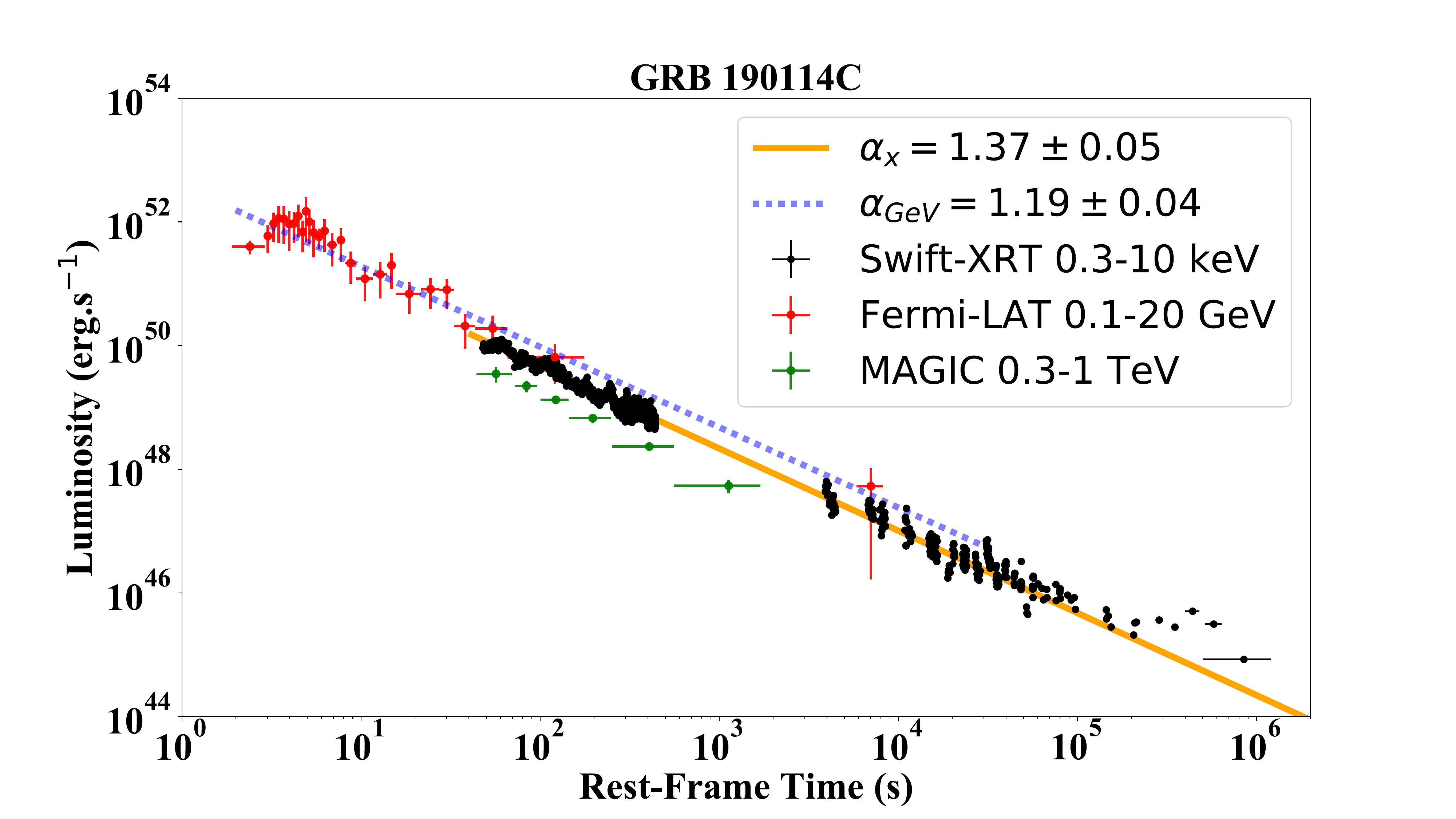} 
\caption{Luminosity of BdHN I 190114C: the black data points represent the rest-frame $0.3$--$10$~keV luminosity obtained from \textit{Swift}-XRT. It follows a decaying power-law with index $\alpha_X=1.37\pm 0.05$. The red data points show the rest-frame $0.1$--$20$~GeV luminosity observed by \textit{Fermi}-LAT. It follows a decaying power-law with amplitude $(4.6\pm 0.6)\times 10^{52}$~erg~s$^{-1}$ and index $\alpha_{\rm GeV}=1.19\pm 0.04$. The green data points show the rest-frame $0.3$--$1$~TeV luminosity obtained from MAGIC. Figure taken from \cite{2021MNRAS.tmp..868R} with the permission of the authors.}\label{fig:GRB190114C}
\end{figure} 

Detailed time-resolved analysis of the lightcurves and spectra of BdHN I have revealed the separated role of the $\nu$NS in the afterglow emission, and of the newborn BH in the GeV emission. The X-ray afterglow observed by the Neil Gehrels \textit{Swift} satellite, characterized by a decreasing luminosity described by a power-law, originates from the synchrotron radiation produced by ultra-relativistic electrons in the expanding SN ejecta, threaded by the magnetic field of the $\nu$NS, and further powered by the $\nu$NS pulsar-like emission \cite{2018ApJ...869..101R, 2019ApJ...874...39W, 2020ApJ...893..148R}. Therefore, the $\nu$NS rotational energy powers the X-ray afterglow emission and we have used this fact to infer from the X-ray data the $\nu$NS spin, as well as the strength and structure of its magnetic field in several sources (see e.g. \cite{2018ApJ...869..101R, 2020ApJ...893..148R}).

The analysis of the GeV emission, characterized by a decreasing luminosity also well fitted by a power-law (but different with respect to the one of the X-rays), explained by the rotational energy extraction from the newborn rotating BH, have allowed to infer for the first time in GRB 130427A \cite{2019ApJ...886...82R}, GRB 190114C \cite{2019arXiv191107552M} and in many other sources in \cite{2021MNRAS.tmp..868R}, the BH mass and spin, as well as the geometrical properties of the GeV emission. The observed lightcurve in the X- and in the high-energy (GeV and beyond) gamma-rays of GRB 190114C is shown in Fig.~\ref{fig:GRB190114C}.

\section{The black hole \textit{inner engine}}\label{sec:3}

We turn now to give qualitative and quantitative details of the BH \textit{inner engine}. The newborn BH in a BdHN I is embedded in the magnetic field inherited from the NS \citep{2020ApJ...893..148R}, and sits at the center of a ``\textit{cavity}'' of very-low density \cite{2019ApJ...883..191R} of material from the HN ejecta (see Fig.~\ref{fig:SPHsimulation}). For GRB 190114C, such a density has been estimated to be of the order of $10^{-14}$~g~cm$^{-3}$. The \textit{cavity} is carved during the accretion and subsequent gravitational collapse of the NS leading to the BH. The magnetic field remains anchored to the material and did not participate in the BH formation (see \cite{2020ApJ...893..148R} for details on the magnetic field surrounding the newborn Kerr BH in a BdHN I).

The Kerr BH in the \textit{cavity} is not isolated, it is surrounded by a magnetic field of strength $B_0$, asymptotically parallel and aligned with the BH rotation axis, and by a fully ionized, very-low-density plasma. The plasma is essential to the electrodynamical performance of the energy extraction process since it feeds the system with the particles to be accelerated. The operation procedure of the BH \textit{inner engine} leads the mass and spin of the BH to be, instead of constant, decreasing functions of time, keeping constant the BH irreducible mass. The electrons accelerate to ultrahigh energy at expenses of the BH rotational energy, and release it via electron-synchrotron photons that carry it off to infinity.

\begin{figure}
    \centering
    \includegraphics[width=\hsize,clip]{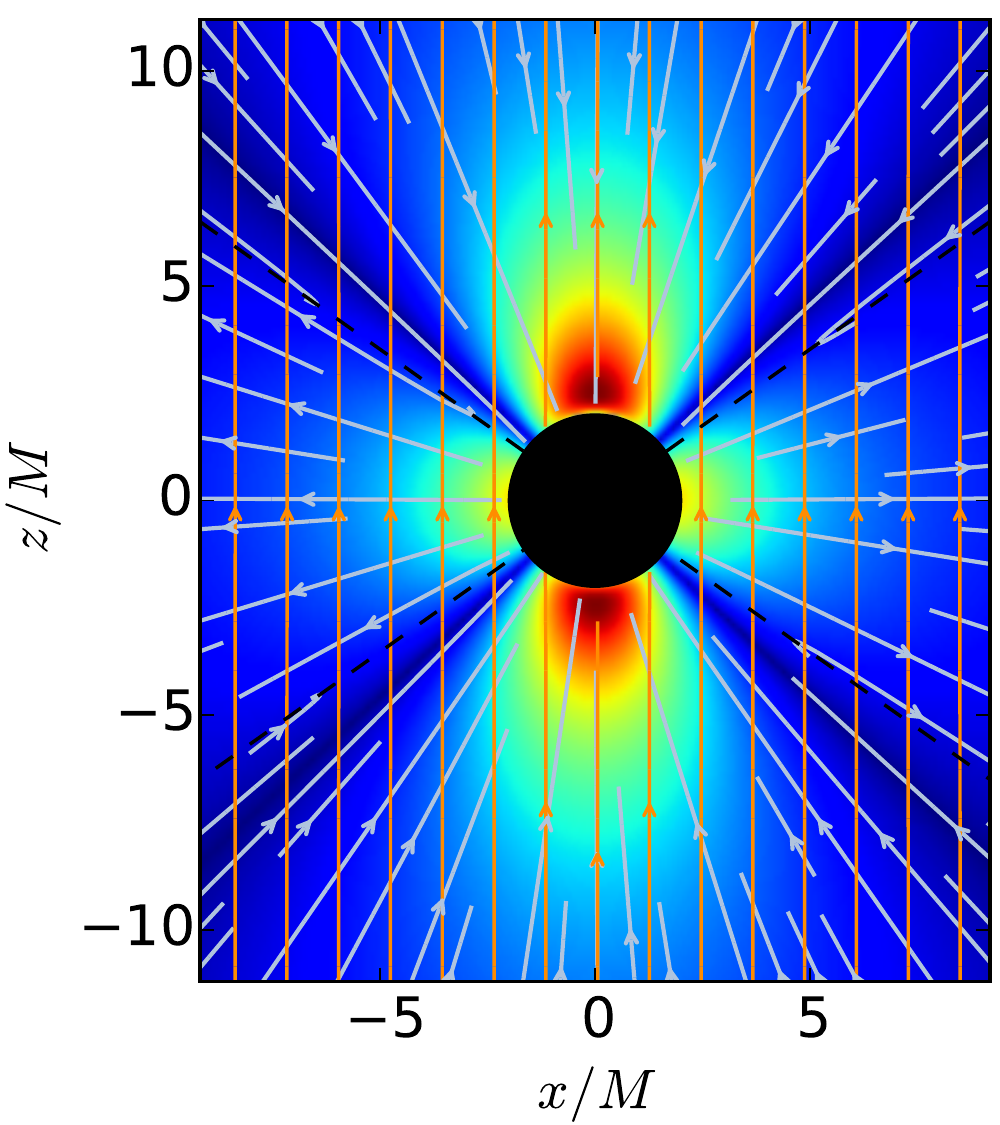}
    \caption{Electric (blue lines) and magnetic (golden lines) field lines of the Papapetrou-Wald solution in the xz plane in Cartesian coordinates. The BH spin parameter is here set to $a/M = 0.3$ and the magnetic field and the BH spin are aligned and parallel. The background is a density-plot of the electric field energy density which is decreasing from red to blue. The BH horizon is the black-filled disk. Distances are in units of $M$ and the fields in units of $B_0$. In the northern hemisphere, the electric field is inwardly-directed in the region covered by spherical polar angles (measured clockwise) $-\theta_\pm<\theta<\theta_\pm$, where $\theta_{\pm} \approx 55^\circ$. By equatorial symmetry, in the southern hemisphere it happens in $\pi-\theta_\pm<\theta<\pi+\theta_\pm$. Electrons located in these northern and southern hemisphere cones of semi-aperture angle of $\approx 60^\circ$ are outwardly accelerated with appropriate pitch angles leading to GeV photons. Clearly, being anisotropic, this ``jetted'' emission is not always visible. This feature is crucial for inferring the morphology of BdHN I from the high-energy (GeV) data of long GRBs \cite{2021MNRAS.tmp..868R}. Figure taken from \cite{2019arXiv191107552M} with permission of the authors.}
    \label{fig:fieldlines}
\end{figure}

A quantitative description of this physical situation can be obtained by means of the solution of the Einstein-Maxwell equations of a Kerr BH embedded in a test, asymptotically aligned, uniform magnetic field \citep{1966AIHPA...4...83P,1974PhRvD..10.1680W}, hereafter the Papapetrou-Wald solution. The BH rotation and the aligned magnetic field induce an electric field that for moderate dimensionless spin values, is mainly radial and inwardly directed. The intensity of this electric field decreases with the square of the distance, has a maximum value at the BH horizon and on the rotation axis ($\theta = 0$), and changes sign at $3 \cos\theta_\pm -1 = 0$. We show in Fig.~\ref{fig:fieldlines} the electric and magnetic field lines in the Papapetrou-Wald solution. The electric field is inwardly-directed in the northern hemisphere for spherical polar angles (measured clockwise) $-\theta_\pm<\theta<\theta_\pm$, where $\theta_{\pm} = \arccos(\sqrt{3}/3) \approx 55^\circ$ (see Fig.~\ref{fig:fieldlines}). Because of the equatorial symmetry, it also points inward in the southern hemisphere for $\pi-\theta_\pm<\theta<\pi+\theta_\pm$. There, electrons are outwardly-accelerated.

The mathematical role of the Papapetrou-Wald solution \citep{1966AIHPA...4...83P, 1974PhRvD..10.1680W} in the BH \textit{inner engine} have led to a profound change of paradigm \citep{2019ApJ...886...82R}, namely the introduction of the effective charge given by the product of $J$ and $B_0$:
\begin{equation}\label{eq:Qeff}
    Q_{\rm eff} = \frac{G}{c^3} 2 J B_0.
\end{equation}
It must be stressed that this charge works as an effective interpretation of the induced electric field which decreases as $1/r^2$ but, actually, the BH is uncharged as it can be shown by integrating the induced surface charge on the whole BH horizon surface. Thus, we are in presence of BHs having ``\textit{charge without charge}'', the electric field arises from the gravitomagnetic interaction of the Kerr BH with the magnetic field. This effective charge, however, allows to finally understand the successful use of a Kerr-Newman BH for the analysis of quantum electrodynamical processes in the field of a rotating BH \cite{1975PhRvL..35..463D}. 

We are now able to elaborate, with the use of quantum electrodynamics and general relativity, a novel and physically more complete treatment of the GRB high-energy engine in a globally neutral system, therefore satisfying Eq.~(\ref{eq:Mbh}) but with $Q=0$!.

\section{The \textit{blackholic quantum}}\label{sec:4}

The operation of the \emph{inner engine} is based on three components naturally present in a BdHN I: (i) the Kerr metric that describes the gravitational field produced by the newborn, rotating BH; (ii) an asymptotically uniform magnetic field around it, fulfilling the Papapetrou-Wald solution (see Fig.~\ref{fig:fieldlines}); (iii) a very-low-density plasma around the newborn BH composed of ions and electrons of $10^{-14}$~g~cm$^{-3}$ \citep{2019ApJ...883..191R}. The BH \textit{inner engine} operates following precise steps:

(1) The gravitomagnetic interaction of the BH spin and the magnetic field induce an electric field as given by the Papapetrou-Wald solution. For an aligned and parallel magnetic field to the BH spin, the electric field is nearly radial and inwardly directed about the BH rotation axis up to an angle $\theta_\pm$ (see Fig.~\ref{fig:fieldlines}). 

(2) The induced electric field accelerates electrons outwardly. The number of electrons that can be accelerated is set by the energy stored in the electric field which, as shown in \cite{2020EPJC...80..300R}, can be expressed in the quantum form:
\begin{equation}
    {\cal E} = \hbar\,\Omega_{\rm eff},
\end{equation}
where $\Omega_{\rm eff}$ is linearly proportional to the BH angular velocity, so depending on the BH mass and spin, and with the proportionality constant depending upon the magnetic field strength, the Planck mass, and the neutron mass. The expression evidences the nature of the underlying physical process generating the electric field and the BH horizon: the electrodynamics of the Papapetrou-Wald solution, the origin of the magnetic field from the NS, and the smooth BH formation from the induced gravitational collapse of the NS by accretion.

(3) The maximum possible electron acceleration/energy is set by the electric potential energy difference from the horizon to infinity \cite{2020EPJC...80..300R}, $\Delta \Phi = e\,a\,B_0/c$.

(4) Along the polar (rotation) axis, radiation losses are absent, therefore electrons can accelerate all the way to reach $\Delta \Phi \approx 10^{18}$~eV, becoming a source of UHECRs. 

(5) At off-axis latitudes, the electrons emit synchrotron radiation responsible of the observed GeV emission.

(6) After this, the energy ${\cal E}$ has been used and emitted. The process restarts with a new angular momentum $J = J_0- \Delta J$, being $\Delta J$ the angular momentum extracted to the Kerr BH by the event. 

The above steps are repeated, with the same efficiency, if the density of plasma is sufficient, namely if the number of the particles is enough to cover the new value of the energy, ${\cal E}$. Therefore, the \emph{inner engine} evolves in a sequence of ``\textit{elementary processes}'', each emitting a well-defined, precise amount of energy ${\cal E}$, the \textit{blackholic quantum}. As an example, we notice that for a magnetic field strength $B_0=10^{11}$~G, BH mass $M=3 M_\odot$ and spin $\alpha= c J/(G M^2) = 0.5$, the \textit{blackholic quantum} of energy is ${\cal E}\approx 3.4\times 10^{37}$~erg, and electrons can be accelerated to energies as large as $\Delta \Phi \approx 10^7$~erg $=6.6\times 10^{18}$~eV!. 

At first sight, one could think this energy can not power the $10^{53}$--$10^{54}$~erg emitted at GeV energies in a long GRB. However, ${\cal E}$ is the energy of each \textit{elementary process}, each \textit{blackholic quantum} which, as we shall see, occurs on timescales as short as $10^{-15}$~s. This leads to luminosities of a few $10^{51}$~erg~s$^{-1}$, just as the one observed (see Fig.~\ref{fig:GRB190114C}). Indeed, in this short timescale only a small fractional angular momentum $\Delta J/J \sim 10^{-16}$ of the Kerr BH is extracted off. Therefore, the process must occur over and over all the way to the resolvable timescales by gamma-ray detectors, e.g. milliseconds and beyond. In fact, a BH angular momentum $\Delta J/J\sim 0.1$ is extracted in the timescale of a few seconds, leading to an extracted energy $E_{\rm extr} \sim 0.1 M c^2$ that explains the observed energy.

\section{Polar and off-polar acceleration}\label{sec:5}

Along the polar axis, $\theta=0$, the electric and magnetic fields are parallel (see Fig.~\ref{fig:fieldlines}). Since the electron is accelerated by the electric field, this implies that the electron pitch angle, i.e. the angle between the electron's injection velocity (into the magnetic field) and the magnetic field is zero. Consequently, no radiation losses (by synchrotron emission) occur for motion along the BH rotation axis. Electrons are accelerated outward along the rotation axis gaining the total electric potential energy, $\Delta \Phi\sim 10^{18}$~eV. Most of this energy is gained at distance scales of the order of the BH horizon, therefore this acceleration occurs on timescales $G M/c^3$ of a few microseconds. This implies that these ejected electrons may contribute to UHECRs at $10^{18}$~eV with a power $\sim 10^{42}$~erg~s$^{-1}$.

\begin{figure}
    \centering
    \includegraphics[width=\hsize,clip]{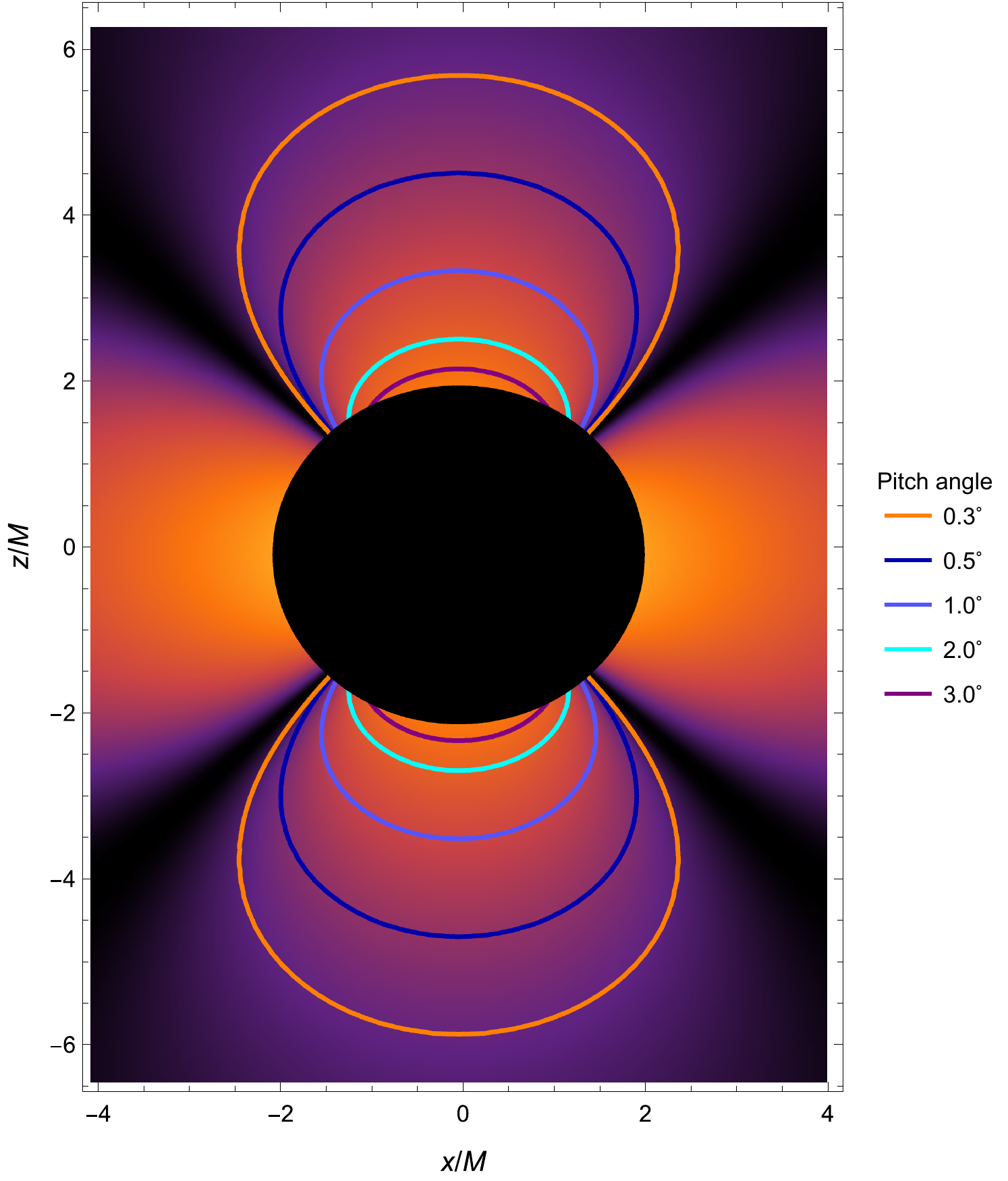}
    \caption{Contours of constant electron pitch angle in the electromagnetic field of the Papapetrou-Wald solution of Fig.~\ref{fig:fieldlines}. The BH is indicated by the filled black disk. The background colormap indicates the electric field energy density (the lighter the more intense). Electrons with these pitch angles emit GeV photons in the approximately conical region with a semi-aperture angle $\theta_\pm \approx 60^\circ$ (dark boundary; see also Fig.~\ref{fig:fieldlines}). This ``\textit{jetted}'' emission is essential to infer the BdHN I morphology from the GeV emission data of long GRBs \cite{2021MNRAS.tmp..868R}. Figure taken from \cite{2019arXiv191107552M} with permission of the authors.}
    \label{fig:pitchangles}
\end{figure}

At off-axis latitudes, the electric and magnetic field cross each other, implying non-zero pitch angles of the accelerated electrons. Figure~\ref{fig:pitchangles} shows contours of constant pitch angle for electrons moving in the electromagnetic field of the Papapetrou-Wald solution shown in Fig.~\ref{fig:fieldlines}.

During the acceleration, the Lorentz factor increases linearly with time up to an asymptotic, maximum value \cite{2019ApJ...886...82R}. This maximum value, set by the balance between the energy gain by acceleration in the electric field, and energy loss by synchrotron radiation. This maximum electron energy leads to photon energies in the GeV regime for the above pitch angles. For instance, in the case of BH spin $\alpha = 0.5$ and $B_0=10^{11}$~G, electrons moving with a pitch angle of $2^\circ$ reach an energy of $2$~GeV, and radiate photons of $1$~GeV (in the energy range of \textit{Fermi}-LAT) with a timescale $3\times 10^{-16}$~s. 

\section{Inferring the BH mass and spin}\label{sec:6}

We require three physical and astrophysical conditions to obtain the three BH \textit{inner engine} parameters, the BH mass and spin, $M$ and $\alpha$, as well as the strength of the magnetic field surrounding the BH, $B_0$. We closely follow the treatment presented in \cite{2019ApJ...886...82R, 2019arXiv191107552M, 2021MNRAS.tmp..868R}. The three conditions are: (1) the GeV energetics is paid by the extractable energy of the BH; (2) the system is transparent to GeV photons produced by the synchrotron radiation of the accelerated electrons; (3) the synchrotron radiation timescale explains the observed GeV emission timescale. This constraint implies that the GeV emission is emitted from electrons being accelerated with appropriate pitch angles (see Fig.~\ref{fig:pitchangles}). Such pitch angles occur within a cone of approximately $60^\circ$ from the BH rotation axis.

From the above, we have inferred for instance for GRB 190114C: $B_0\approx 3.9\times 10^{10}$~G, spin and BH mass, respectively, $\alpha = 0.41$ and $M=4.45~M_\odot$. The BH irreducible mass is $M_{\rm irr}=4.35~M_\odot$. We have applied a minimum energy requirement, namely that $E_{\rm extr}$ equals the observed energy at GeV energies, so these values are lower limits. The BH \textit{inner engine} can be extremely long-lasting, it can continue to radiate more than the observed ($10^4$~s of emission) $E_{\rm GeV}$. Because of the power-law behavior of the GeV luminosity, most of the energy is emitted in this early evolution, and just a small higher value of mass and spin can make the system to work for much longer times. For example, let us assume that this observed power-law luminosity (see Fig.~\ref{fig:GRB190114C}) extends to longer times, we can check that a BH with $\alpha = 0.46$, $M=4.50~M_\odot$, and $M_{\rm irr}=4.38~M_\odot$ can power a $25\%$  larger GeV emission energy, keeping the BH radiating for $1000$~yr!. 

In \cite{2021MNRAS.tmp..868R}, we applied this method to several BdHN I and inferred BH masses $2.3$--$8.9~M_\odot$ and spin $0.27$--$0.87$. Besides explaining the GeV emission from the BH energy extraction, the time evolution validates, time by time, the BH mass-energy formula.

\section{Conclusions}\label{sec:7}

The \textit{inner engine} uses an efficient gravito-electrodynamical process that explains the GeV emission of long GRBs. The gravito-magnetic interaction of the Kerr BH spin with the surrounding magnetic field induces an electric field which accelerates electrons from the BH vicinity. The kinetic energy gained by electrons is radiated off to infinity by synchrotron emission due to the presence of the magnetic field.

It is worth stressing that there is not any bulk motion: each electron is accelerated to a maximum energy set by the balance between electric acceleration and synchrotron radiation losses. The  electron-synchrotron photons have energies in the GeV domain. The radiation of the BH \textit{inner engine} e.g. at keV to MeV energies is negligible (with respect to the observed values). The observed radiation in the keV to MeV energy domains is explained by a different mechanism in a BdHN I; see e.g. \cite{2020ApJ...893..148R}. 
The request that the observed GeV emission be paid by the extractable (rotational) energy of the Kerr BH have allowed us to estimate, for the first time, the mass and spin of BHs in long GRBs. Since we have here used only the GeV observational data, these values of the BH mass and spin must be considered as lower limits. In fact, we have shown that even a small higher mass (or spin) of the BH can guarantee even larger and longer emission of the BH \textit{inner engine}, and in view of the decaying power-law behavior of the GeV emission, it may lasts forever!. 

Before closing, it is worth to recall some crucial aspects of the BH \textit{inner engine}. (I) The nature of the emission results from considering the physical process leading to the electric and magnetic fields and the BH formation. (II) This is fundamental to show that the emission process leading to the observed luminosity is not continuous but discrete. (III) The timescale of the emission in GRBs is too short to be probed directly by current observational facilities. Direct evidence of the value and discreteness might come out instead from the observation of large BHs of $10^8$--$10^{10}~M_\odot$ in AGN. For instance, in the case of M87*, for $M = 6\times 10^9~M_\odot$, $\alpha = 0.1$, and $B_0=10~G$, the BH \textit{inner engine} theory predicts a high-energy (GeV) emission with a luminosity of a few $10^{43}$~erg~s$^{-1}$, with a timescale of up to tenths of seconds, while the timescale for UHECRs emission is of the order of half a day.

All the above results are important. The underlying proof that indeed we can use the extractable rotational energy of a Kerr BH to explain the high-energy \textit{jetted} emissions of GRBs and AGN stands alone. The \textit{jetted} emission does not originate from ultra-relativistic acceleration of matter in bulk (massive jets), but from very special energy-saving general relativistic and electrodynamical processes leading to the emission of \emph{blackholic quanta} of energy \cite{2020EPJC...80..300R}. We were waiting for this result for fifty years since ``Introducing the black hole'' \cite{Ruffini:1971bza} and the writing of equation (\ref{eq:Mbh}). We are happy to have given the evidence of the successful operation of the BH \textit{inner engine} in this 50th anniversary.


\begin{thebibliography}{25}
\expandafter\ifx\csname natexlab\endcsname\relax\def\natexlab#1{#1}\fi
\expandafter\ifx\csname bibnamefont\endcsname\relax
  \def\bibnamefont#1{#1}\fi
\expandafter\ifx\csname bibfnamefont\endcsname\relax
  \def\bibfnamefont#1{#1}\fi
\expandafter\ifx\csname citenamefont\endcsname\relax
  \def\citenamefont#1{#1}\fi
\expandafter\ifx\csname url\endcsname\relax
  \def\url#1{\texttt{#1}}\fi
\expandafter\ifx\csname urlprefix\endcsname\relax\def\urlprefix{URL }\fi
\providecommand{\bibinfo}[2]{#2}
\providecommand{\eprint}[2][]{\url{#2}}

\bibitem[{\citenamefont{{Kerr}}(1963)}]{1963PhRvL..11..237K}
\bibinfo{author}{\bibfnamefont{R.~P.} \bibnamefont{{Kerr}}},
  \bibinfo{journal}{\prl} \textbf{\bibinfo{volume}{11}}, \bibinfo{pages}{237}
  (\bibinfo{year}{1963}).

\bibitem[{\citenamefont{{Newman} et~al.}(1965)\citenamefont{{Newman}, {Couch},
  {Chinnapared}, {Exton}, {Prakash}, and {Torrence}}}]{1965JMP.....6..918N}
\bibinfo{author}{\bibfnamefont{E.~T.} \bibnamefont{{Newman}}},
  \bibinfo{author}{\bibfnamefont{E.}~\bibnamefont{{Couch}}},
  \bibinfo{author}{\bibfnamefont{K.}~\bibnamefont{{Chinnapared}}},
  \bibinfo{author}{\bibfnamefont{A.}~\bibnamefont{{Exton}}},
  \bibinfo{author}{\bibfnamefont{A.}~\bibnamefont{{Prakash}}},
  \bibnamefont{and}
  \bibinfo{author}{\bibfnamefont{R.}~\bibnamefont{{Torrence}}},
  \bibinfo{journal}{Journal of Mathematical Physics}
  \textbf{\bibinfo{volume}{6}}, \bibinfo{pages}{918} (\bibinfo{year}{1965}).

\bibitem[{\citenamefont{{Christodoulou}}(1970)}]{1970PhRvL..25.1596C}
\bibinfo{author}{\bibfnamefont{D.}~\bibnamefont{{Christodoulou}}},
  \bibinfo{journal}{\prl} \textbf{\bibinfo{volume}{25}}, \bibinfo{pages}{1596}
  (\bibinfo{year}{1970}).

\bibitem[{\citenamefont{{Christodoulou} and
  {Ruffini}}(1971)}]{1971PhRvD...4.3552C}
\bibinfo{author}{\bibfnamefont{D.}~\bibnamefont{{Christodoulou}}}
  \bibnamefont{and}
  \bibinfo{author}{\bibfnamefont{R.}~\bibnamefont{{Ruffini}}},
  \bibinfo{journal}{\prd} \textbf{\bibinfo{volume}{4}}, \bibinfo{pages}{3552}
  (\bibinfo{year}{1971}).

\bibitem[{\citenamefont{{Hawking}}(1972)}]{1972CMaPh..25..152H}
\bibinfo{author}{\bibfnamefont{S.~W.} \bibnamefont{{Hawking}}},
  \bibinfo{journal}{Communications in Mathematical Physics}
  \textbf{\bibinfo{volume}{25}}, \bibinfo{pages}{152} (\bibinfo{year}{1972}).

\bibitem[{\citenamefont{Ruffini and Wheeler}(1971)}]{Ruffini:1971bza}
\bibinfo{author}{\bibfnamefont{R.}~\bibnamefont{Ruffini}} \bibnamefont{and}
  \bibinfo{author}{\bibfnamefont{J.~A.} \bibnamefont{Wheeler}},
  \bibinfo{journal}{Phys. Today} \textbf{\bibinfo{volume}{24}},
  \bibinfo{pages}{30} (\bibinfo{year}{1971}).

\bibitem[{\citenamefont{DeWitt and DeWitt}(1973)}]{DeWitt:1973uma}
\bibinfo{editor}{\bibfnamefont{C.}~\bibnamefont{DeWitt}} \bibnamefont{and}
  \bibinfo{editor}{\bibfnamefont{B.~S.} \bibnamefont{DeWitt}}, eds.,
  \emph{\bibinfo{title}{{Proceedings, Ecole d'EtÃ© de Physique ThÃ©orique: Les
  Astres Occlus}}}, vol.~\bibinfo{volume}{23} of \emph{\bibinfo{series}{Les
  Houches Summer School}}, \bibinfo{organization}{Gordon and Breach}
  (\bibinfo{publisher}{Gordon and Breach}, \bibinfo{address}{New York, NY},
  \bibinfo{year}{1973}).

\bibitem[{\citenamefont{{Penrose} and {Floyd}}(1971)}]{1971NPhS..229..177P}
\bibinfo{author}{\bibfnamefont{R.}~\bibnamefont{{Penrose}}} \bibnamefont{and}
  \bibinfo{author}{\bibfnamefont{R.~M.} \bibnamefont{{Floyd}}},
  \bibinfo{journal}{Nature Physical Science} \textbf{\bibinfo{volume}{229}},
  \bibinfo{pages}{177} (\bibinfo{year}{1971}).

\bibitem[{\citenamefont{{Ruffini}
  et~al.}(2019{\natexlab{a}})\citenamefont{{Ruffini}, {Moradi}, {Rueda},
  {Becerra}, {Bianco}, {Cherubini}, {Filippi}, {Chen}, {Karlica}, {Sahakyan}
  et~al.}}]{2019ApJ...886...82R}
\bibinfo{author}{\bibfnamefont{R.}~\bibnamefont{{Ruffini}}},
  \bibinfo{author}{\bibfnamefont{R.}~\bibnamefont{{Moradi}}},
  \bibinfo{author}{\bibfnamefont{J.~A.} \bibnamefont{{Rueda}}},
  \bibinfo{author}{\bibfnamefont{L.}~\bibnamefont{{Becerra}}},
  \bibinfo{author}{\bibfnamefont{C.~L.} \bibnamefont{{Bianco}}},
  \bibinfo{author}{\bibfnamefont{C.}~\bibnamefont{{Cherubini}}},
  \bibinfo{author}{\bibfnamefont{S.}~\bibnamefont{{Filippi}}},
  \bibinfo{author}{\bibfnamefont{Y.~C.} \bibnamefont{{Chen}}},
  \bibinfo{author}{\bibfnamefont{M.}~\bibnamefont{{Karlica}}},
  \bibinfo{author}{\bibfnamefont{N.}~\bibnamefont{{Sahakyan}}},
  \bibnamefont{et~al.}, \bibinfo{journal}{\apj} \textbf{\bibinfo{volume}{886}},
  \bibinfo{eid}{82} (\bibinfo{year}{2019}{\natexlab{a}}), \eprint{1812.00354}.

\bibitem[{\citenamefont{{Rueda} and {Ruffini}}(2020)}]{2020EPJC...80..300R}
\bibinfo{author}{\bibfnamefont{J.~A.} \bibnamefont{{Rueda}}} \bibnamefont{and}
  \bibinfo{author}{\bibfnamefont{R.}~\bibnamefont{{Ruffini}}},
  \bibinfo{journal}{European Physical Journal C} \textbf{\bibinfo{volume}{80}},
  \bibinfo{eid}{300} (\bibinfo{year}{2020}), \eprint{1907.08066}.

\bibitem[{\citenamefont{{Becerra} et~al.}(2019)\citenamefont{{Becerra},
  {Ellinger}, {Fryer}, {Rueda}, and {Ruffini}}}]{2019ApJ...871...14B}
\bibinfo{author}{\bibfnamefont{L.}~\bibnamefont{{Becerra}}},
  \bibinfo{author}{\bibfnamefont{C.~L.} \bibnamefont{{Ellinger}}},
  \bibinfo{author}{\bibfnamefont{C.~L.} \bibnamefont{{Fryer}}},
  \bibinfo{author}{\bibfnamefont{J.~A.} \bibnamefont{{Rueda}}},
  \bibnamefont{and}
  \bibinfo{author}{\bibfnamefont{R.}~\bibnamefont{{Ruffini}}},
  \bibinfo{journal}{\apj} \textbf{\bibinfo{volume}{871}}, \bibinfo{eid}{14}
  (\bibinfo{year}{2019}), \eprint{1803.04356}.

\bibitem[{\citenamefont{{Rueda} and {Ruffini}}(2012)}]{2012ApJ...758L...7R}
\bibinfo{author}{\bibfnamefont{J.~A.} \bibnamefont{{Rueda}}} \bibnamefont{and}
  \bibinfo{author}{\bibfnamefont{R.}~\bibnamefont{{Ruffini}}},
  \bibinfo{journal}{\apjl} \textbf{\bibinfo{volume}{758}}, \bibinfo{eid}{L7}
  (\bibinfo{year}{2012}), \eprint{1206.1684}.

\bibitem[{\citenamefont{{Fryer} et~al.}(2014)\citenamefont{{Fryer}, {Rueda},
  and {Ruffini}}}]{2014ApJ...793L..36F}
\bibinfo{author}{\bibfnamefont{C.~L.} \bibnamefont{{Fryer}}},
  \bibinfo{author}{\bibfnamefont{J.~A.} \bibnamefont{{Rueda}}},
  \bibnamefont{and}
  \bibinfo{author}{\bibfnamefont{R.}~\bibnamefont{{Ruffini}}},
  \bibinfo{journal}{\apjl} \textbf{\bibinfo{volume}{793}}, \bibinfo{eid}{L36}
  (\bibinfo{year}{2014}), \eprint{1409.1473}.

\bibitem[{\citenamefont{{Becerra} et~al.}(2016)\citenamefont{{Becerra},
  {Bianco}, {Fryer}, {Rueda}, and {Ruffini}}}]{2016ApJ...833..107B}
\bibinfo{author}{\bibfnamefont{L.}~\bibnamefont{{Becerra}}},
  \bibinfo{author}{\bibfnamefont{C.~L.} \bibnamefont{{Bianco}}},
  \bibinfo{author}{\bibfnamefont{C.~L.} \bibnamefont{{Fryer}}},
  \bibinfo{author}{\bibfnamefont{J.~A.} \bibnamefont{{Rueda}}},
  \bibnamefont{and}
  \bibinfo{author}{\bibfnamefont{R.}~\bibnamefont{{Ruffini}}},
  \bibinfo{journal}{\apj} \textbf{\bibinfo{volume}{833}}, \bibinfo{eid}{107}
  (\bibinfo{year}{2016}), \eprint{1606.02523}.

\bibitem[{\citenamefont{{Becerra} et~al.}(2015)\citenamefont{{Becerra},
  {Cipolletta}, {Fryer}, {Rueda}, and {Ruffini}}}]{2015ApJ...812..100B}
\bibinfo{author}{\bibfnamefont{L.}~\bibnamefont{{Becerra}}},
  \bibinfo{author}{\bibfnamefont{F.}~\bibnamefont{{Cipolletta}}},
  \bibinfo{author}{\bibfnamefont{C.~L.} \bibnamefont{{Fryer}}},
  \bibinfo{author}{\bibfnamefont{J.~A.} \bibnamefont{{Rueda}}},
  \bibnamefont{and}
  \bibinfo{author}{\bibfnamefont{R.}~\bibnamefont{{Ruffini}}},
  \bibinfo{journal}{\apj} \textbf{\bibinfo{volume}{812}}, \bibinfo{eid}{100}
  (\bibinfo{year}{2015}), \eprint{1505.07580}.

\bibitem[{\citenamefont{{Fryer} et~al.}(2015)\citenamefont{{Fryer}, {Oliveira},
  {Rueda}, and {Ruffini}}}]{2015PhRvL.115w1102F}
\bibinfo{author}{\bibfnamefont{C.~L.} \bibnamefont{{Fryer}}},
  \bibinfo{author}{\bibfnamefont{F.~G.} \bibnamefont{{Oliveira}}},
  \bibinfo{author}{\bibfnamefont{J.~A.} \bibnamefont{{Rueda}}},
  \bibnamefont{and}
  \bibinfo{author}{\bibfnamefont{R.}~\bibnamefont{{Ruffini}}},
  \bibinfo{journal}{Physical Review Letters} \textbf{\bibinfo{volume}{115}},
  \bibinfo{eid}{231102} (\bibinfo{year}{2015}), \eprint{1505.02809}.

\bibitem[{\citenamefont{{Ruffini} et~al.}(2021)\citenamefont{{Ruffini},
  {Moradi}, {Rueda}, {Li}, {Sahakyan}, {Chen}, {Wang}, {Aimuratov}, {Becerra},
  {Bianco} et~al.}}]{2021MNRAS.tmp..868R}
\bibinfo{author}{\bibfnamefont{R.}~\bibnamefont{{Ruffini}}},
  \bibinfo{author}{\bibfnamefont{R.}~\bibnamefont{{Moradi}}},
  \bibinfo{author}{\bibfnamefont{J.~A.} \bibnamefont{{Rueda}}},
  \bibinfo{author}{\bibfnamefont{L.}~\bibnamefont{{Li}}},
  \bibinfo{author}{\bibfnamefont{N.}~\bibnamefont{{Sahakyan}}},
  \bibinfo{author}{\bibfnamefont{Y.~C.} \bibnamefont{{Chen}}},
  \bibinfo{author}{\bibfnamefont{Y.}~\bibnamefont{{Wang}}},
  \bibinfo{author}{\bibfnamefont{Y.}~\bibnamefont{{Aimuratov}}},
  \bibinfo{author}{\bibfnamefont{L.}~\bibnamefont{{Becerra}}},
  \bibinfo{author}{\bibfnamefont{C.~L.} \bibnamefont{{Bianco}}},
  \bibnamefont{et~al.}, \bibinfo{journal}{\mnras}  (\bibinfo{year}{2021}),
  \eprint{2103.09142}.

\bibitem[{\citenamefont{{Wang} et~al.}(2019)\citenamefont{{Wang}, {Rueda},
  {Ruffini}, {Becerra}, {Bianco}, {Becerra}, {Li}, and
  {Karlica}}}]{2019ApJ...874...39W}
\bibinfo{author}{\bibfnamefont{Y.}~\bibnamefont{{Wang}}},
  \bibinfo{author}{\bibfnamefont{J.~A.} \bibnamefont{{Rueda}}},
  \bibinfo{author}{\bibfnamefont{R.}~\bibnamefont{{Ruffini}}},
  \bibinfo{author}{\bibfnamefont{L.}~\bibnamefont{{Becerra}}},
  \bibinfo{author}{\bibfnamefont{C.}~\bibnamefont{{Bianco}}},
  \bibinfo{author}{\bibfnamefont{L.}~\bibnamefont{{Becerra}}},
  \bibinfo{author}{\bibfnamefont{L.}~\bibnamefont{{Li}}}, \bibnamefont{and}
  \bibinfo{author}{\bibfnamefont{M.}~\bibnamefont{{Karlica}}},
  \bibinfo{journal}{\apj} \textbf{\bibinfo{volume}{874}}, \bibinfo{eid}{39}
  (\bibinfo{year}{2019}), \eprint{1811.05433}.

\bibitem[{\citenamefont{{Ruffini} et~al.}(2018)\citenamefont{{Ruffini},
  {Karlica}, {Sahakyan}, {Rueda}, {Wang}, {Mathews}, {Bianco}, and
  {Muccino}}}]{2018ApJ...869..101R}
\bibinfo{author}{\bibfnamefont{R.}~\bibnamefont{{Ruffini}}},
  \bibinfo{author}{\bibfnamefont{M.}~\bibnamefont{{Karlica}}},
  \bibinfo{author}{\bibfnamefont{N.}~\bibnamefont{{Sahakyan}}},
  \bibinfo{author}{\bibfnamefont{J.~A.} \bibnamefont{{Rueda}}},
  \bibinfo{author}{\bibfnamefont{Y.}~\bibnamefont{{Wang}}},
  \bibinfo{author}{\bibfnamefont{G.~J.} \bibnamefont{{Mathews}}},
  \bibinfo{author}{\bibfnamefont{C.~L.} \bibnamefont{{Bianco}}},
  \bibnamefont{and}
  \bibinfo{author}{\bibfnamefont{M.}~\bibnamefont{{Muccino}}},
  \bibinfo{journal}{\apj} \textbf{\bibinfo{volume}{869}}, \bibinfo{eid}{101}
  (\bibinfo{year}{2018}), \eprint{1712.05000}.

\bibitem[{\citenamefont{{Rueda} et~al.}(2020)\citenamefont{{Rueda}, {Ruffini},
  {Karlica}, {Moradi}, and {Wang}}}]{2020ApJ...893..148R}
\bibinfo{author}{\bibfnamefont{J.~A.} \bibnamefont{{Rueda}}},
  \bibinfo{author}{\bibfnamefont{R.}~\bibnamefont{{Ruffini}}},
  \bibinfo{author}{\bibfnamefont{M.}~\bibnamefont{{Karlica}}},
  \bibinfo{author}{\bibfnamefont{R.}~\bibnamefont{{Moradi}}}, \bibnamefont{and}
  \bibinfo{author}{\bibfnamefont{Y.}~\bibnamefont{{Wang}}},
  \bibinfo{journal}{\apj} \textbf{\bibinfo{volume}{893}}, \bibinfo{eid}{148}
  (\bibinfo{year}{2020}), \eprint{1905.11339}.

\bibitem[{\citenamefont{{Moradi} et~al.}(2019)\citenamefont{{Moradi}, {Rueda},
  {Ruffini}, and {Wang}}}]{2019arXiv191107552M}
\bibinfo{author}{\bibfnamefont{R.}~\bibnamefont{{Moradi}}},
  \bibinfo{author}{\bibfnamefont{J.~A.} \bibnamefont{{Rueda}}},
  \bibinfo{author}{\bibfnamefont{R.}~\bibnamefont{{Ruffini}}},
  \bibnamefont{and} \bibinfo{author}{\bibfnamefont{Y.}~\bibnamefont{{Wang}}},
  \bibinfo{journal}{arXiv e-prints} \bibinfo{eid}{arXiv:1911.07552}
  (\bibinfo{year}{2019}), \eprint{1911.07552}.

\bibitem[{\citenamefont{{Ruffini}
  et~al.}(2019{\natexlab{b}})\citenamefont{{Ruffini}, {Melon Fuksman}, and
  {Vereshchagin}}}]{2019ApJ...883..191R}
\bibinfo{author}{\bibfnamefont{R.}~\bibnamefont{{Ruffini}}},
  \bibinfo{author}{\bibfnamefont{J.~D.} \bibnamefont{{Melon Fuksman}}},
  \bibnamefont{and} \bibinfo{author}{\bibfnamefont{G.~V.}
  \bibnamefont{{Vereshchagin}}}, \bibinfo{journal}{\apj}
  \textbf{\bibinfo{volume}{883}}, \bibinfo{eid}{191}
  (\bibinfo{year}{2019}{\natexlab{b}}).

\bibitem[{\citenamefont{{Papapetrou}}(1966)}]{1966AIHPA...4...83P}
\bibinfo{author}{\bibfnamefont{A.}~\bibnamefont{{Papapetrou}}},
  \bibinfo{journal}{Annales de L'Institut Henri Poincare Section (A) Physique
  Theorique} \textbf{\bibinfo{volume}{4}}, \bibinfo{pages}{83}
  (\bibinfo{year}{1966}).

\bibitem[{\citenamefont{{Wald}}(1974)}]{1974PhRvD..10.1680W}
\bibinfo{author}{\bibfnamefont{R.~M.} \bibnamefont{{Wald}}},
  \bibinfo{journal}{\prd} \textbf{\bibinfo{volume}{10}}, \bibinfo{pages}{1680}
  (\bibinfo{year}{1974}).

\bibitem[{\citenamefont{{Damour} and {Ruffini}}(1975)}]{1975PhRvL..35..463D}
\bibinfo{author}{\bibfnamefont{T.}~\bibnamefont{{Damour}}} \bibnamefont{and}
  \bibinfo{author}{\bibfnamefont{R.}~\bibnamefont{{Ruffini}}},
  \bibinfo{journal}{\prl} \textbf{\bibinfo{volume}{35}}, \bibinfo{pages}{463}
  (\bibinfo{year}{1975}).

\end{thebibliography}
\end{document}